\documentclass{elsarticle}

\biboptions{authoryear}

\usepackage{graphicx}
\usepackage{multirow,booktabs,array}
\usepackage{amsmath,amsfonts,amssymb,mathrsfs,bm}
\usepackage[hyperfigures=true,colorlinks=true,citecolor=blue,linkcolor=blue]{hyperref}

\begin{document}
%
%
\begin{frontmatter}

\title{Sand transverse dune aerodynamics:\\
3D Coherent Flow Structures from a computational study
}

\author[label2,label4]{Luca Bruno\corref{cor1}}\ead{luca.bruno@polito.it}\ead[url]{http://www.polito.it/wsmm}
 \author[label3,label4]{Davide Fransos}

\cortext[cor1]{Corresponding author. Tel: (+39) 011.090.4870. Fax: (+39) 011.090.4999.}
\address[label2]{Politecnico di Torino, Department of Architecture and Design,\\ Viale Mattioli 39, I-10125, Torino, Italy}
\address[label3]{Optiflow Company, \\ 160, Chemin de la Madrague-Ville, F-13015, Marseille, France}
\address[label4]{Windblown Sand Modeling and Mitigation joint research group}

\begin{abstract}
The engineering interest about dune fields is dictated by the their interaction with a number of human infrastructures in arid environments.
Sand dunes dynamics is dictated by wind and its ability to induce sand erosion, transport and deposition.
A deep understanding of dune aerodynamics serves then to ground effective strategies for the protection of human infrastructures from sand, the so-called sand mitigation. 
Because of their simple geometry and their frequent occurrence in desert area, transverse sand dunes are usually adopted in literature as a benchmark to investigate dune aerodynamics by means of both computational or experimental approaches, usually in nominally 2D setups.
The present study aims at evaluating 3D flow features in the wake of a idealised transverse dune, if any, under different nominally 2D setup conditions by means of computational simulations and to compare the obtained results with experimental measurements available in literature.
\end{abstract}


\end{frontmatter}
%
%
%
%
\section{Introduction}\label{sec:intro}

The engineering interest about dune fields is dictated by their interaction with a number of human infrastructures in arid environments, such as roads and railways, pipelines, industrial facilities, farms, buildings \citep[e.g.][]{Alghamdi:2005em}.
{Some of such undesired effects are shown in Figure \ref{fig:sand_challenges}.
\begin{figure}[h]
\begin{center}
\includegraphics[width=\textwidth]{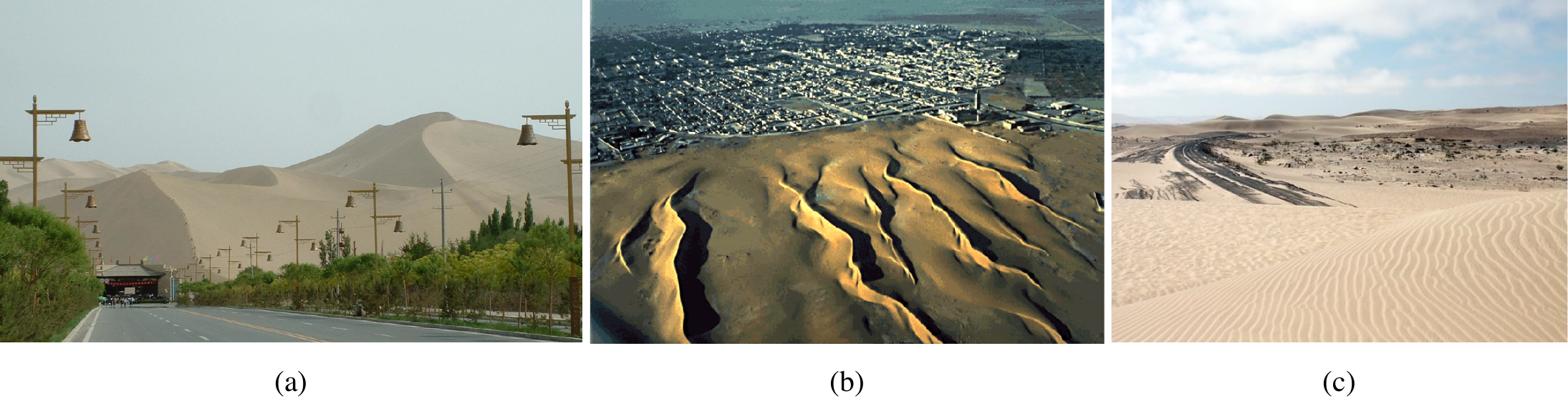}
\caption{Windblown sand interaction with anthropic activities: megadunes surrounding a road and farmlands in Dunhuang, Gansu Province, PRC. Photo: I.A. Inman, 2007 (a), linear dunes encroaching Nouakchott, capital of Mauritania. Landsat 1565-10032-6, 1974 (b), loose sand covering the Aus to L\"{u}deritz railway line, Namibia. Photo: K. Dierks, 2003 (c)}
\label{fig:sand_challenges}
\end{center}
\end{figure}
}
The development, shape and migration of dunes depend fundamentally on availability of mobile sediment, on the incoming wind directionality, and  on the flow structures of local disturbed wind ("topographically forced wind" in Geomorphology literature), that is on the wind ability to induce sand erosion, transport and deposition.
In particular, coherent flow structures embedded within fluid flow fields are considered to govern the magnitude, form and scaling of sediment transport events \citep{Bauer:2013gd}, and the dune shape in turn. Hence, a wind-focused perspective has been adopted in aeolian dune geomorphology in order to investigate the former to explain the latter.
In wind engineering, a deep understanding of dune aerodynamics serves to ground effective strategies for the protection of human infrastructures from sand, the so-called sand mitigation.\\
In areas of constant wind direction and under high sand availability, the transverse dune - which has nearly fixed profile in the direction perpendicular to the wind - is the prevailing dune type \citep[][]{Livingstone:1996ve}. Furthermore, ideal sharp-crested transverse dune, i.e. without crest-brink separation \citep{Bauer:2013gd}, are usually adopted in fundamental aerodynamic studies for their geometric simplicity. Previous studies \citep[e.g.][]{Lancaster:1995wy} have revealed that transverse dunes generally have upwind (windward, stoss) slope angles ranging between $2^{\circ}\leq\alpha_u\leq 20^{\circ}$ and downwind (leeward) slope angles ranging between $28^{\circ}\leq\alpha_d\leq 34^{\circ}$, i.e. around the sand friction static angle.
For sake of clarity, a simplified 2D scheme of the transverse dune geometry and of the flow around it is given in Figure \ref{fig:aerod-scheme}.
\begin{figure}[htbp]
\begin{center}
\includegraphics[]{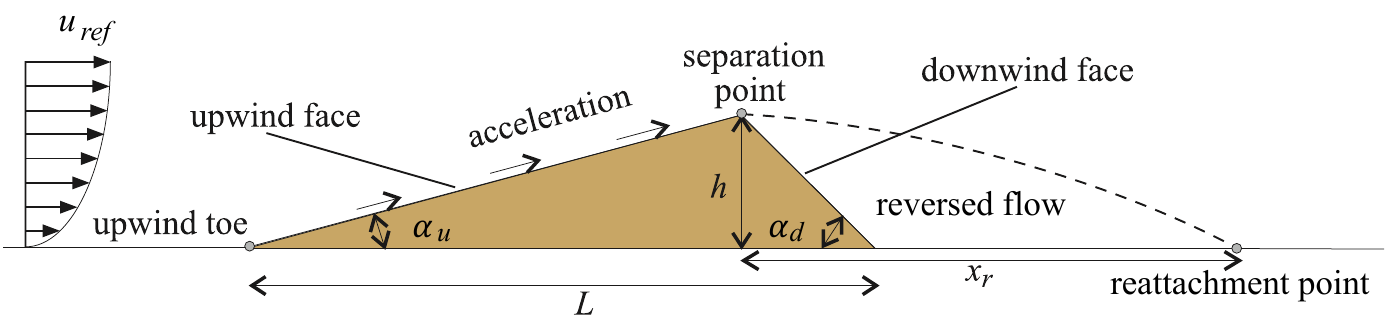}
\caption{Ideal sharp crest, transverse dune aerodynamics scheme}
\label{fig:aerod-scheme}
\end{center}
\end{figure}
The aerodynamic behaviour of sand dunes in atmospheric boundary layer belongs to the very general class of bluff bodies 
{and to the general one of  hills \citep[for a review, see e.g.][]{Bitsuamlak_2004}.}
The incoming wind flow is first fairly decelerated at the upwind toe, then strongly accelerated along the dune upwind face straight up to crest. The separation of the boundary layer then occurs at the crest itself. The naturally inclined downwind face is surrounded by a reversed flow region, whose extent is one of the key parameters describing the flow. Reattachment of the boundary layer occurs far downstream the dune crest, being $x_r$ the so-called reattachment length. The latter quantity is experienced in sharp-edge bluff body aerodynamics to be highly sensitive to a number of setup parameters (e.g. Reynolds number, surface roughness, incoming turbulence).
Much more detailed 2D topological mapping of the wake region for isolated transverse dunes under crest-normal flow can be recovered in literature \citep[e.g.][]{Walker:2002hw,Walker:2003dd,DelgadoFernandez:2011}.
\\
%
%
%
While the 2D flow structures above have been extensively scrutinized and reviewed in \cite{Livingstone:2007if}, very little is known about three-dimensional coherent flow structures in the wake \citep[for an updated review, see][]{Bauer:2013gd}.
On one hand, in the Geomorphology community, such 3D coherent flow structures past nominally 2D dunes are usually ascribed to: i. oblique incoming wind, i.e. where a yaw angle $\alpha_C > 45^{\circ}$ gives rise to helical vortices in the wake \citep{Allen_1970,Walker:2002hw}, and/or ii. small crest irregularities, inducing spanwise flow in the wake, spanwise swirling of the reversed flow, and secondary recirculation zones \citep[see the recent CFD simulations and field measurements in ][, respectively]{Jackson_2011, Jackson_2013, DelgadoFernandez:2013dt}. It is worth pointing out that from an aerodynamic point of view both occurrences involve a 3D setup.
On the other hand, to the writers best knowledge, 3D coherent structures in the wake of genuine 2D setups (i.e. ideal transverse dune under normal wind) remain elusive and scarcely studied in literature, if any.
Difficulties in both wind tunnel facilities and measurement techniques, and in 3D computational simulations are conjectured by the writers and other authors \citep[e.g.][]{Walker:2003dd} to be the cause of such lack of knowledge.
\\
On one side, Wind Tunnel (WT) tests in the aeolian field adopt scaled dune models spanning across the whole test section width (i.e. $S=w$), and with rather low values of the dune aspect ratio (span/chord ratio $S/L$), e.g. $S/L= 1.28$ in \cite{Walker:2000wc} and \cite{Walker:2003dd}, $1.92 \leq S/L \leq  10.32$ in \cite{Dong:2007gm,Qian:2009dc}, $S/L= 5.4$ in \cite{Liu:2011iha}). Such values of the aspect ratios:
\begin{itemize}
\item are by far lower than the ones adopted in other bluff body aerodynamics problems characterised by separation, reversed flow and reattachment, especially when 3D flow features are expected \citep[see for instance][]{bruno_2014};
\item are suspected to significantly affect the 3D features of the flow in the wake, because of the tips effects due to interaction between the boundary layers surrounding the dune and the WT side walls. For instance, in \citet[][Figure 5a, page 1119]{Walker:2003dd} the normalised shear stress profile along the WT midline (i.e. .46 m far from the WT side wall) does not reach nil value in the conjectured reattachment area nor elsewhere in the dune wake.
\end{itemize}
More generally, difficulties in WT studies are recognised in the proper scaling of different setup lengths, i.e. the ones related to the dune geometry, the surface roughness, the turbulent length scale of the turbulent incoming flow \citep{Walker:2003dd}.
\\
On the other side, the fundamental computational (CFD) studies on the flow field over the transverse dune \citep[e.g.][]{Parsons:2004gm,Parsons:2004haa,Schatz:2006iy,Araujo:2013gy} usually consider simplified 2D condition \citep[see][for a review]{Livingstone:2007if}. Only recently, \cite{Liu:2011iha} have compared the simulated flow around a transverse dune obtained in 2D and "2.5D" conditions between them and with WT measurements. The "2.5D" heading indicates that despite the computational domain is 3D, its height and width are equal to half the WT working section dimensions where reference experimental test are performed. In other terms, both left-to-right and bottom-to-top symmetry of the flow are conjectured to reduce the domain size and related computational costs.
Under such an assumption, interesting preliminary results are obtained. In particular, 3D flow features of the flow in the near wake of the 2D dune are qualitatively observed \citep[][Figure 8a, page 884]{Liu:2011iha}.
According to the writers, some questions immediately follow. Which aerodynamic phenomena underly and/or trigger the 3D flow features observed in the cited experiments and computational simulations? In particular, do the WT side wall effects play any role in generating such 3D structures? That is, would similar structures having corresponding characteristic lengths obtain if the same dune had, in the limit, an infinitely long span (i.e., an infinite aspect ratio)?
\\
\indent
The present study aims at shedding some light on such issues. The sensitivity of the 3D features of the wake to the setup geometrical scaling is studied. In particular, the domain size and the related inlet and side boundary conditions
are the retained parameters of the study.
In such a perspective, the present study takes the baton relay from the stimulating study of \citep{Liu:2011iha} conceived in the Geomorphology community, and develops it according to the knowledge background of Bluff Body Aerodynamics.
The adopted computational approach is expected to efficiently complement the wind tunnel studies in exploring a huge number of setup conditions, where a single parameter is varied at the time and border, or even unphysical, conditions can be scrutinized.
The experimental setup of \cite{Liu:2011iha} is adopted as the reference one. The obtained computational results are compared with the experimental ones available in literature. Finally, a deeper insight in the 3D emerging coherent {flow} structures in the wake is provided.

\section{Wind flow modelling and computational approach}
\label{sec:modelling}
%
%
%
The incompressible, turbulent, separated, unsteady flow around the dune profile is modeled by the classical Time-dependent Reynolds
Averaged Navier-Stokes (T-RANS) equations, which, in Cartesian coordinates, read:
\begin{equation}
\dfrac{\partial \overline{u_i}}{\partial x_i}=0
\end{equation}
\begin{equation}
\dfrac{\partial \overline{u_i}}{\partial t}+\overline{u_j}\dfrac{\partial \overline{u_i}}{\partial x_j}=
-\dfrac{1}{\rho}\dfrac{\partial \overline{p}}{\partial x_i}+\dfrac{\partial }{\partial x_j}\Big[\nu\Big( \dfrac{\partial \overline{u_i}}{\partial x_j}
+\dfrac{\partial \overline{u_j}}{\partial x_i} \Big)\Big]-\dfrac{\partial}{\partial x_j}(\overline{u'_iu'_j}),
\end{equation}
where $\overline{u_i}$ is the averaged velocity, $u'$ the velocity fluctuating component, $\overline{p}$ the averaged pressure,
$\rho$ the air density and $\nu$ the air kinematic viscosity.
{
The SST $k-\omega$ turbulence model first proposed by 
\cite{Menter:1994kl} and further modified in \cite{Menter:2003wm} is used to close the T-RANS equations:
}
\begin{equation}
\frac {\partial k}{\partial t}+ \overline{u}_i\, \frac {\partial k}{\partial x_i
}=\frac {\partial }{\partial x_i }\left[\Big(\sigma_k \nu_t+\nu\Big)\,
\frac {\partial k}{\partial x_i }\right]+\tilde{P}_k-\beta^{\ast}k\omega
\end{equation}
\begin{equation}
\frac {\partial \omega}{\partial t}+ \overline{u}_i\, \frac {\partial \omega}{\partial x_i
}=\frac {\partial }{\partial x_i }\left[\Big(\sigma_{\omega}\nu_t+\nu\Big)\,
\frac {\partial \omega}{\partial x_i }\right]+\alpha\dfrac{\omega}{k}P_k-\beta\omega^{2} + \left( 1-F_1\right) \frac{2\sigma_{\omega}}{\omega}\frac {\partial k}{\partial x_i }\frac {\partial \omega}{\partial x_i },
\end{equation}
where $k$ is the turbulent kinetic energy, $\omega$ its specific dissipation rate and $\nu_t$ the 
so-called turbulent kinematic viscosity.
The kinetic energy production term $\tilde{P}_{k}$ is modeled by introducing a production limiter to prevent the build-up of turbulence in stagnation regions:
\[\tilde{P}_{k}=\min\left( P_k, 10\beta^{\ast} k \omega \right) \quad \textrm{being} \quad P_k\approx 2\nu_t D_{ij}\dfrac{\partial \overline{u}_i}{\partial x_i}.\]
For sake of conciseness, the definition of the blending function $F_1$ and the values of the model constants are omitted herein. Interested readers can find them in \cite{Menter:2003wm}.
{
The SST $k-\omega$ turbulence model is selected for the current application because of its proven accuracy in bluff body aerodynamics in general \citep{Menter:2003wm} and in dune aerodynamics in particular \citep{Liu:2011iha}.
\\
}
%
%
{
At the ground and dune surfaces the so-called sand-grain roughness wall functions are selected for the current application because of their wide use in environmental CWE in general \citep[e.g.][]{Blocken:2007b} and the proofs of adequacy obtained in previous 3D simulations of sand dune aerodynamics by \cite{Liu:2011iha, Jackson_2011, Jackson_2013}.
In particular, standard wall functions \citep{Launder:1974} with roughness modification \citep{Cebeci:1977} are applied.
The required equivalent sand grain roughness height is determined as $K_s=9.793z_0/C_s$, where $C_s=0.5$ is the roughness constant, and the first cell at wall is limited by $n_p>K_S$, being $n_p$ half of the cell size, following \cite{Blocken:2007jq}.
}
\\
\indent
The adopted computational domains and the conditions imposed at their boundaries are the object of a parametrical study. They are detailed in the next Section.
\\
%
%
%
\indent
The OpenFoam\copyright Finite Volume open source code is used in the following to numerically evaluate the flow-field.
The cell-centre values of the variables are interpolated at face locations using the second-order Central Difference Scheme for the diffusive terms.
The convection terms are discretised by means of the so-called Limited Linear scheme, a 2nd order accurate bounded Total Variational Diminishing (TVD) scheme resulting from the application of the Sweby limiter \citep{Sweby1984} to the central differencing in order to enforce a monotonicity criterion.
The pressure-velocity coupling is achieved by means of the pressure-implicit PISO
algorithm, using a predictor-corrector approach for the time discretisation of the momentum equation,
whilst enforcing the continuity equation.
{
The space discretization is accomplished by a predominantly structured grid of hexahedral control volumes. Unstructured patterns locally occur at the intersection between the surface-fitted grid boundary layer and the cartesian grid in the higher part of the domain.
The denser grid is located close to the wind tunnel floor and side walls, and to the dune surface. The height of the control volume adjacent to the wall $n_w$ is driven by the sand-grain roughness wall function requirements \citep{Blocken:2007jq}. In particular, $n_w$ is set equal to $n_w= 0.2 h$ in order to satisfy at best both a sufficiently high mesh resolution in the normal direction $n$ to the surface, and the requirement $n_p = n_w/2 > K_S$, being for the adopted mesh $n_p / K_S = 1.28$.
}
The advancement in time is accomplished by the implicit two-step second order Backward Differentiation Formulae (BDF) method.
The adopted time step is equal to $\Delta t = 0.00025$ [s], i.e. $\Delta t u_{ref}/h= 0.1$ dimensionless time unit.
%
%
%
\section{Application setup}\label{sec:appl_setup}
\begin{figure}[htbp]
\begin{center}
\includegraphics[width=\textwidth]{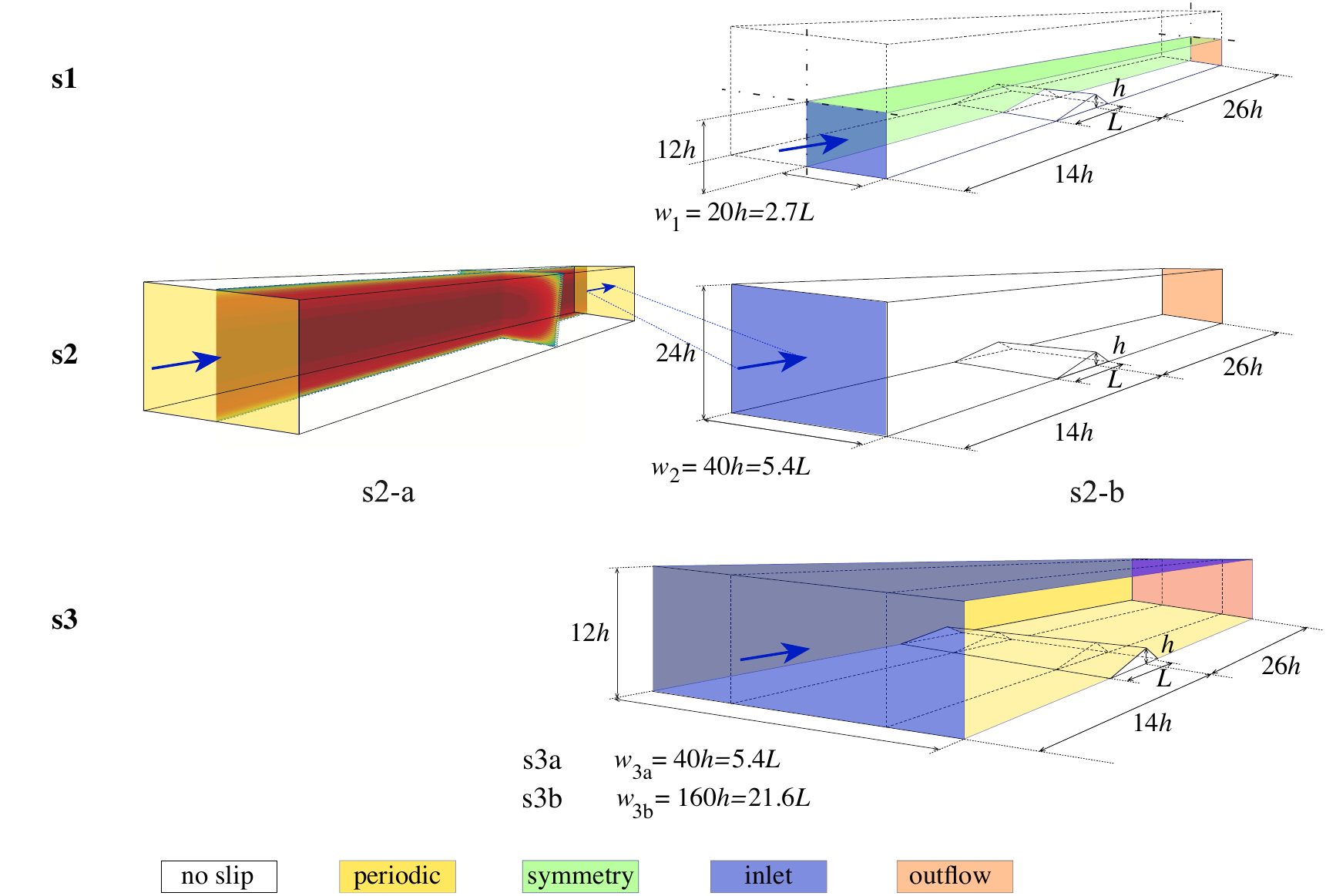}
\caption{Scheme of the adopted setup conditions (not in scale) including b.c. (wind from left to right)}
\label{fig:adopted-setup}
\end{center}
\end{figure}

\begin{figure}[htbp]
\begin{center}
\includegraphics[width=\textwidth]{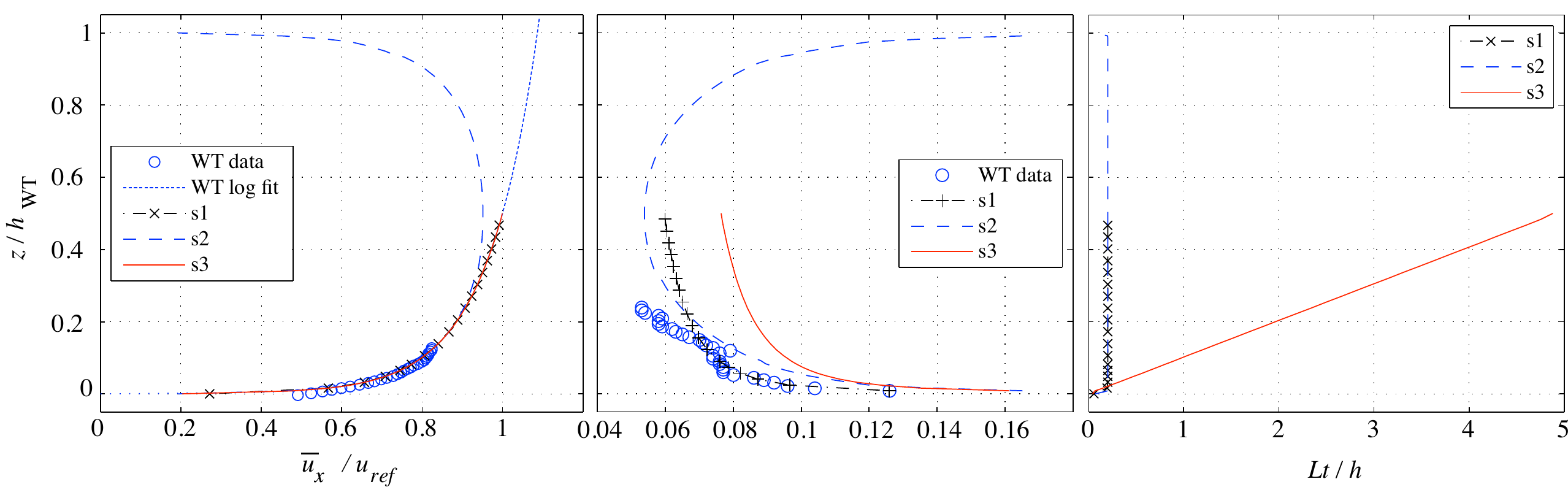}
\caption{Vertical profiles of the inlet boundary conditions}
\label{fig:adopted-inlet-bc}
\end{center}
\end{figure}

%
%
{
Three setups are adopted and the resulting flow are compared.
All of them adopt a sharp-crested transverse dune, and generally refer to the experimental setup described by \cite{Liu:2011iha}.
Besides such main reference, critical comparison is also made to other measurements provided in \cite{Dong:2007gm,Qian:2009dc,Walker:2003dd}, 
the former two being obtained for several incoming speeds, the latter in slightly different experimental conditions. Table \ref{tab:setups} summarizes the main features of the cited experimental setups, where $z_0$ is the dune and floor aerodynamic roughness and $u_{ref}$ is the incoming reference speed.
\begin{table}[h!]
\caption{Experimental setups}
\begin{center}
\begin{tabular}{lccccccc}
Authors 			& $h$	& $\alpha_u$ 	& $\alpha_d$ 	& $L/h$ 	& $S/L$ 	& $u_{ref}$ 	& $z_0$ 	\\
				& [mm] 	& [deg] 		& [deg] 		& [-]		& [-]		& [m/s] 		& [mm]	\\
\hline
\cite{Liu:2011iha} 	&  		&  			&  			&  		& 	 	& 10 			& 		\\
\cite{Dong:2007gm} 	& 25 		& 10 			& 30 			& 7.4 	& 5.40 	& 8,10,12,14 	& 0.1		\\
\cite{Qian:2009dc}	&  		&  			&  			&	 	& 	 	& 8,10,12,14	& 		\\
\cite{Walker:2003dd}	& 80 		& 8   			& 30 			& 9.0 	& 1.28 	& 8,13,18 		& n.a.	\\
\hline
\end{tabular}
\end{center}
\label{tab:setups}
\end{table}
It follows that the reference Reynolds number is equal to $Re_h=u_{ref} h / \nu=1.7e+4$. 
}
\\
The setups differ in both the size of the analytical domain in space and in the corresponding applied boundary conditions (b.c.).
The domain size and the kind of b.c. adopted in the setups are depicted in Figure \ref{fig:adopted-setup}, while the profiles of the incoming wind (mean velocity $u_x$, turbulence intensity $It$ and length scale $Lt$) are plotted in Figure \ref{fig:adopted-inlet-bc}. The setups are briefly commented in the following:
\begin{itemize}
\item[\bf s1] only 1/4 of the wind tunnel working section volume is retained. Symmetry conditions are imposed on the vertical plane because of the conjectured symmetry of the flow. Free stream b.c.s are set on the upper horizontal plane to disregard the influence of the WT upper wall. The incoming mean wind velocity and turbulence intensity profiles are fitted on the WT measurements \citep[][, estimated shear velocity $u^*=0.512$ m/s]{Liu:2011iha}, while turbulence length scale is conjectured constant and equal to $Lt$ = 5 mm because of the lack of experimental data (Figure \ref{fig:adopted-inlet-bc}).
{
The spatial grid involves about $4.e+5$ control volumes.
}
This setup is often retained in CFD practice to reduce the number of grid volumes and related computational costs when the WT conditions are to be emulated \citep[e.g. in][]{Liu:2011iha}. The setup is intended to discuss the accuracy of such an approach;
\item[\bf s2] the setup exactly reproduces the size of the WT working section in \cite{Dong:2007gm,Qian:2009dc,Liu:2011iha}. No-slip b.c.s are set at the four alongwind boundary planes to model the WT walls.
A complementary simulation (s2-a, Figure \ref{fig:adopted-setup}) is preliminary performed along an empty channel to replicate the WT approaching section described in \cite{Liu:2011iha}. Periodic conditions at the inlet and outlet of s2-a are set to obtain a {fully self-developed, horizontally homogeneous incoming boundary layer flow \citep{Blocken:2007jq}}. The resulting $u_x$, $It$ and $Lt$ profiles are then imposed at the inlet of the main domain (Figure \ref{fig:adopted-inlet-bc}).
{
The spatial grid in s2-b involves about $1.5e+6$ control volumes.
}
The setup aims at removing the ansatz introduced in s1 in order to suggest best practice guidelines in CFD simulations of WT tests;
\item[\bf s3] The side walls of the WT working section are replaced by spanwise periodic conditions.
{Two domain spanwise lengths are considered: $w_{3a}=w_2$ is set in setup s3a, and $w_{3a}=4w_2$ in setup s3b}.
The height of the domain is set equal to $12h$, i.e. larger than the range suggested in \cite{Franke:2007tp}  $4h<h_{WT}<10h$ for external flows in Wind Engineering.
{
The spatial grid in s3b involves about $2.7e+6$ control volumes.
}
The setup aims at comparing the results to the ones obtained in the s2 setup and at evaluating 3D flow features, if any, under 2D, "external" incoming flow conditions.
The profiles of the incoming turbulent length scale and turbulence intensity are set in accordance to \cite{Richards_2011} to replicate an external flow.
\end{itemize}
In all setups, Neumann conditions ("outflow" in Figure \ref{fig:adopted-setup}) involving the velocity field and the pressure (null normal component of the stress tensor) as well as $k$ and $\omega$ are imposed at the outlet boundary. No-slip conditions are imposed at the floor surface and at the WT walls, if any.
Uniform initial conditions are imposed at the beginning of the time-dependent simulations.
The simulated time for each setup is equal to about $T u_{ref}/h= 4000$ dimensionless time units, long enough to guarantee the convergence of the first and second statistical moments in time for all the flow variable, following the convergence check proposed by \cite{Bruno_2010}.
{
The simulations have been performed thanks to the Optiflow pc cluster (32 nodes with 4 cores each, 4 Gb RAM per node, Intel Nehalem 2.8 GHz clock, 1.4 Tflops).
An overall cpu time of about {25} hours on 63 cores results for the complete parametrical study.
}
%
\section{Results}\label{sec:resu}
%
\subsection{Setup effects on the overall flow regime}\label{sec:resu_1}
\begin{figure}[htbp]
\begin{center}
\includegraphics[width=\textwidth]{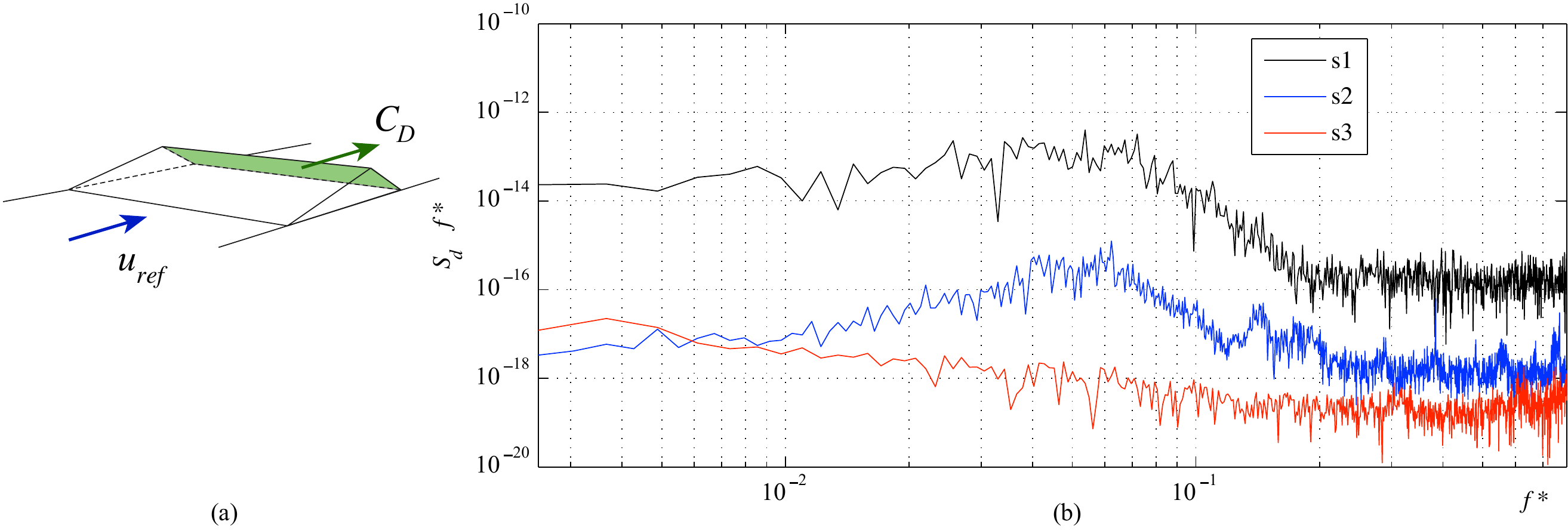}
\caption{Time dependent flow: PSD of the drag coefficient}
\label{fig:drag-psd}
\end{center}
\end{figure}

In this section, the time dependent behavior of the simulated flow in the three setups is discussed with reference to both bulk and local quantities.
The former is defined as the aerodynamic coefficient $C_D(t)$ of drag force {per spanwise unit length} that results from integration of the stress field on the dune downwind face, Figure \ref{fig:drag-psd}(a).
The Power Spectral Density of the drag coefficient is plotted versus the dimensionless frequency $f^*=fh/u_{ref}$ in Figure \ref{fig:drag-psd}(b).
The PSDs in both s1 and s2 setups clearly show a well defined peak at about $f^* \approx 0.05 = St$, that conversely disappears in {s3a and s3b setups (both notated "s3" in the following, if not specified otherwise, for the sake of simplicity)}.
At such Strouhal number $St$ the power density in setup s1 is one order of magnitude higher than the one in s2, that is in turn by far greater than in s3. The high frequency fluctuations ($f^* > 3e-1$) are due to spurious numerical oscillations of the solution.
In summary, the drag force resulting from s3 is almost steady ($std(C_D)\approx 5.6e-7$), while very weak to moderate time fluctuations { in the remaining setups ( $std(C_D)\approx 7.3e-5$  and $std(C_D)\approx 1.e-3$ in s2 and s1, respectively )} suggest a localized unsteadiness of the flow.
\begin{figure}[h]
\begin{center}
\includegraphics[width=\textwidth]{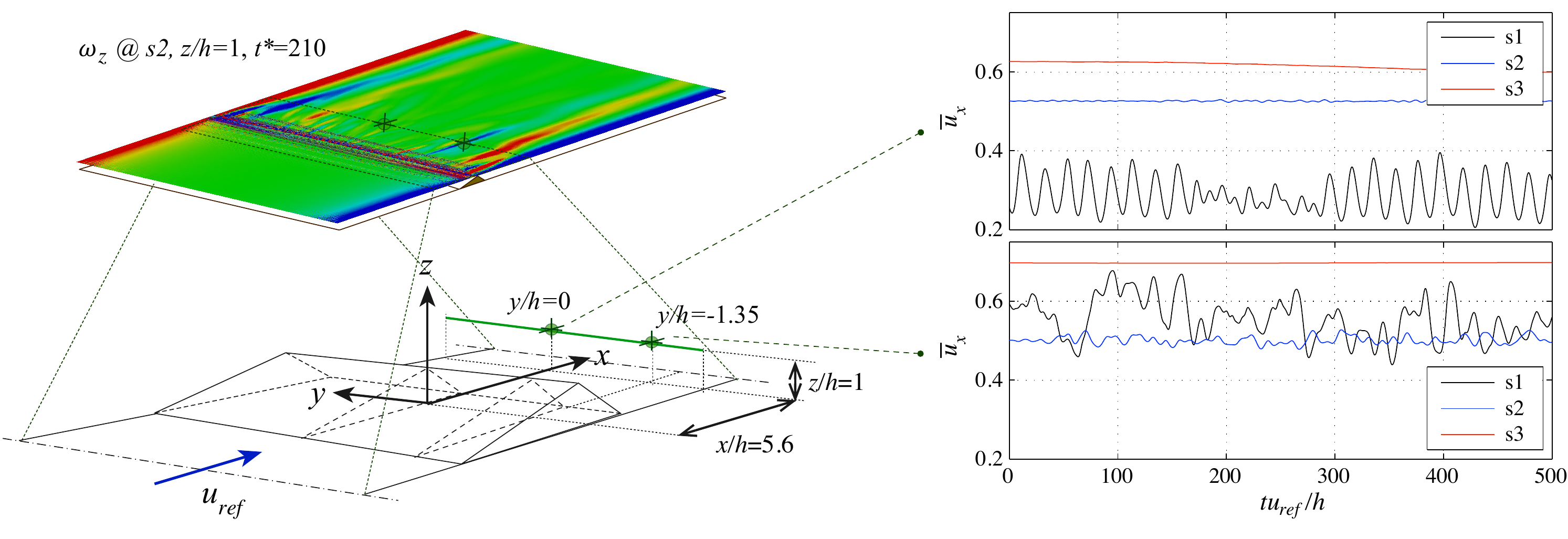}
\caption{Time dependent flow: spanwise position of the time dependent phenomena}
\label{fig:dune-wake-probes}
\end{center}
\end{figure}
To prove such conjecture, the time history of the longitudinal component of the velocity $u_x$ is plotted for the three setups and at two probes in the wake (Figure \ref{fig:dune-wake-probes}). 
In s1 and s2, the closer the probe to the WT side wall, the larger the velocity fluctuations, while in s3 the flow in the wake is almost steady everywhere.
An instantaneous field of the $z$-vorticity on the plane $z/h=1$ for s2 is provided in the same Figure to shed some light on the underlying fluid flow phenomenon. In fact, the interaction between the separated boundary layer at the dune crest and the attached one along the WT sidewall induces a 3D flapping along the latter.
\\
As previously introduced, the reattachment length $x_r$ is one of the most relevant quantities in transverse dune aerodynamics. In the adopted computational approach, the reattachment point is evaluated as the one in which the longitudinal component $\tau_x$ of the shear stress change sign downwind the dune downwind toe. Heaving in ming the weak variability in time of the overall flow, the time averaged reattachment point ($t-avg(x_r)$) is evaluated, while its spanwise trend is described by the reattachment line, i.e. $t-avg(x_r(y))$.
The latter ones obtained in the three setups are plotted in Figure \ref{fig:reattachment}.
\\
\begin{figure}[h]
\begin{center}
\includegraphics[width=\textwidth]{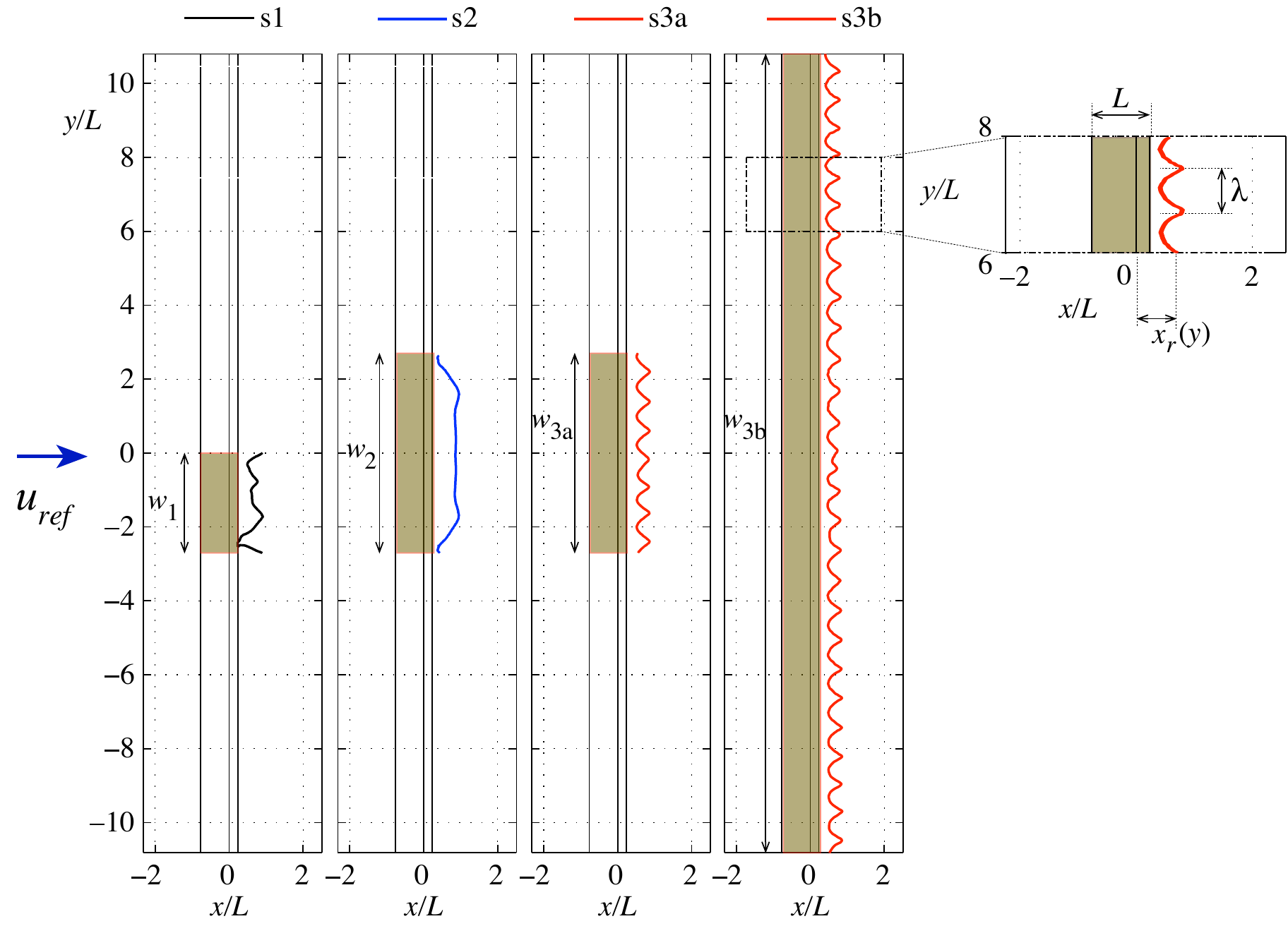}
\caption{Spanwise dependent flow: variation of the reattachment line}
\label{fig:reattachment}
\end{center}
\end{figure}
The following remarks follow:
\begin{itemize}
\item very significant WT wall effects take place close to the dune tips in both s1 and s2 setups. In particular:
\begin{itemize}
\item [-] in s1 such effects propagate along all the dune span, so that a meaningful estimate of the reattachment length can be obtained by spanwise averaging nor by adopting a point wise value, e.g. at the mid plane;
\item [-] in s2 the spanwise variability due to the WT side wall is limited to a distance from the wall of about $d \approx 1.5L$, clearly related to the unsteady plumes pointed out in Figure \ref{fig:dune-wake-probes}. A central segment with nearly constant $x_r$ can be recognized: at the vertical mid plan ($y=0$) $x_r \approx 6.35 h$, pretty close to the value measured by \cite{Dong:2007gm} ($x_r \approx 6 h$) in the same setup conditions;
\end{itemize}
\item unexpectedly, a quasi-periodic, "festoon-shaped" spanwise reattachment line emerges from {both simulations s3a and s3b}. Such a trend allows to estimate some spanwise statistics.
{
\begin{itemize}
\item [-] 
The first two statistical moments of the reattachment length, i.e. its mean value $y-avg(t-avg(x_r))$  and standard deviation $y-std(t-avg(x_r))$, are:\\
{\bf{s3a}}  $y-avg (t-avg ( x_r )) = 5.04 h$,   $y-std (t-avg ( x_r )) = 0.82 h$;\\
{\bf{s3b}}  $y-avg (t-avg ( x_r )) = 4.64 h$,   $y-std (t-avg ( x_r )) = 0.88 h$.\\
It is worth pointing out that the in both simulations the mean value $y-avg (t-avg ( x_r ))$ is significantly lower than the one obtained at the mid-plan in s2.
\item [-] 
Let us define a characteristic spanwise length (or wavelength) $\lambda$ of the reattachment line (see closeup view in Figure \ref{fig:reattachment}). Its spanwise statistics are:\\
{\bf{s3a}}  $y-avg (t-avg ( \lambda )) = 5.61 h$,   $y-std (t-avg ( \lambda )) = 0.43 h$;\\
{\bf{s3b}}  $y-avg (t-avg ( \lambda )) = 5.75 h$,   $y-std (t-avg ( \lambda )) = 0.36 h$.\\
The value of the standard deviation $y-std (t-avg ( \lambda ))$ is small in both simulations, that is the festoon shape is quite stable spanwise.
\item [-] 
Finally it is worth pointing out that the spanwise dimension $w_{3a}$ in setup s3a accommodates 7 wavelengths (including the two halves at the domain side), while the one $w_{s3b}=4 w_{s3a}$ in setup s3b accommodates 27 waves plus an oscillation at midspan, i.e. at the far section from the lateral periodic b.c.s.
\end{itemize}
}
\end{itemize}
In summary, the findings obtained in the Section prove significant effects of the WT side walls (s1 and s2) on the overall flow regime, and suggest that they inhibit the triggering of an almost periodic spanwise variability of the flow in the wake, as observed in external flow conditions (s3).
\subsection{Comparison between WT measurements and CWE results}\label{sec:resu_2}
In this Section, comparisons between time-averaged WT measurements at the dune midspan ($y/L=0$) and the present computational results obtained in the different setups are provided. Bearing in mind the quasi-periodic spanwise reattachment line in s3 (Fig. \ref{fig:reattachment}), also spanwise statistics are evaluated for such setup.
%
\begin{figure}[h!]
\begin{center}
\includegraphics[width=\textwidth]{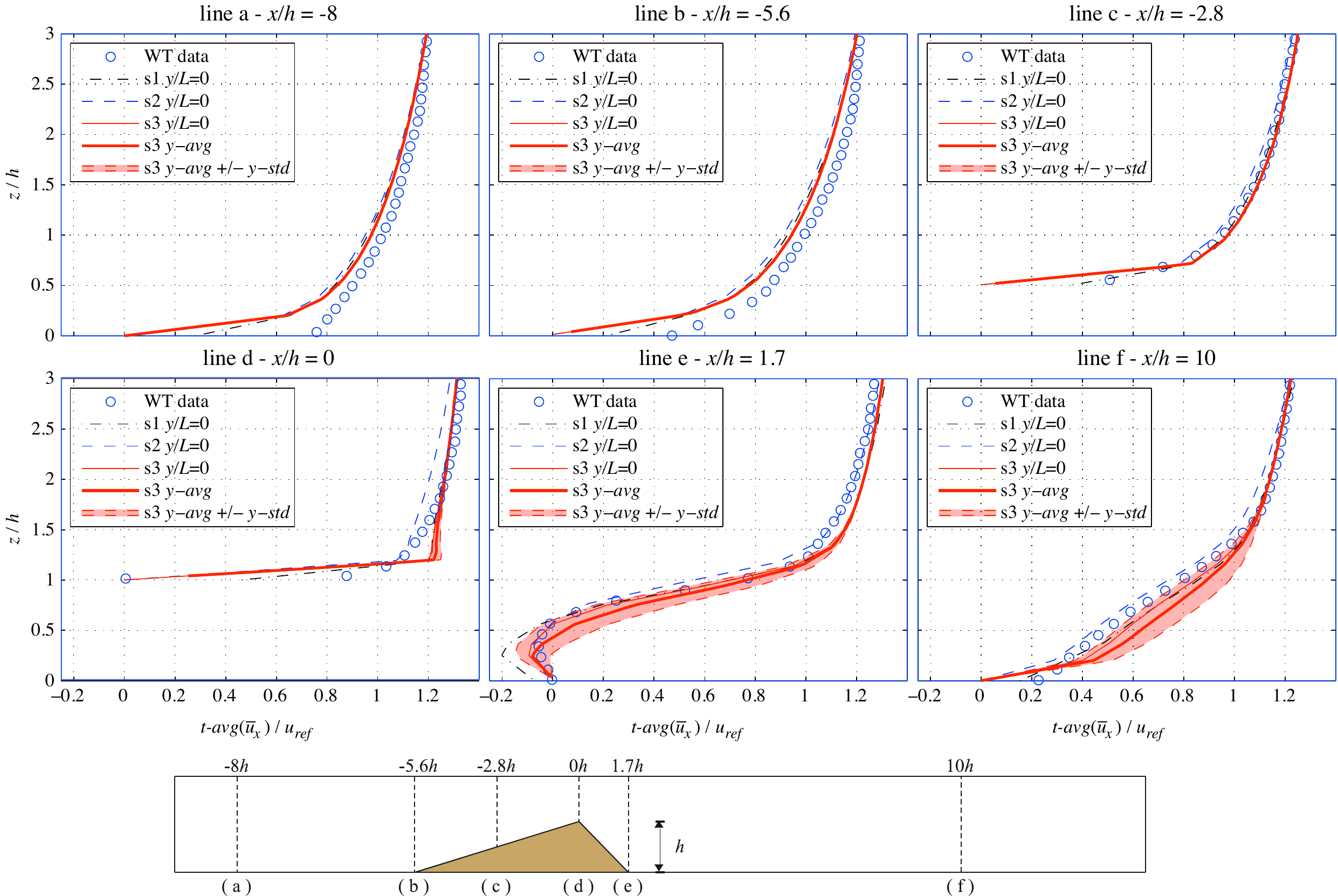}
\caption{Horizontal velocity profiles: WT measurements \citep{Liu:2011iha} and present CWE results}
\label{fig:ux_profiles}
\end{center}
\end{figure}
The $z$-wise profiles of the longitudinal component of the velocity $u_x$ have been measured in WT tests by \cite{Liu:2011iha} using PIV at different positions in the the vertical mid plan ($y=0$) of the WT working section. These profiles are compared to the computational ones in Figure \ref{fig:ux_profiles}.
The scatter with the WT data around the upwind face (Figure \ref{fig:ux_profiles}, lines a-b) are probably due, according to \cite{Liu:2011iha}, to the limitation of the PIV system, which cannot fully resolve the high-speed gradient in the near-surface zone and leads to nonzero wind speed at the surface. The agreement in the reversed flow region (Figure \ref{fig:ux_profiles}, lines e-f) is excellent for setup s2, that replicates the WT conditions. The computational results from setup s1 strongly overestimate the speed deficit in the near wake, while the $y-avg(t-avg(\overline u_x))$ in s3 is quite different form the centre plane one and from WT data. In the same setup, a significant spanwise deviation occurs in the wake (Figure \ref{fig:ux_profiles}, lines e-f).
\\
The measured $x$-wise profiles of the vertical component of the velocity $u_y$ at the level $z \approx h$ are reported in \cite{Qian:2009dc}. 
\begin{figure}[h!]
\begin{center}
\includegraphics[width=0.73\textwidth]{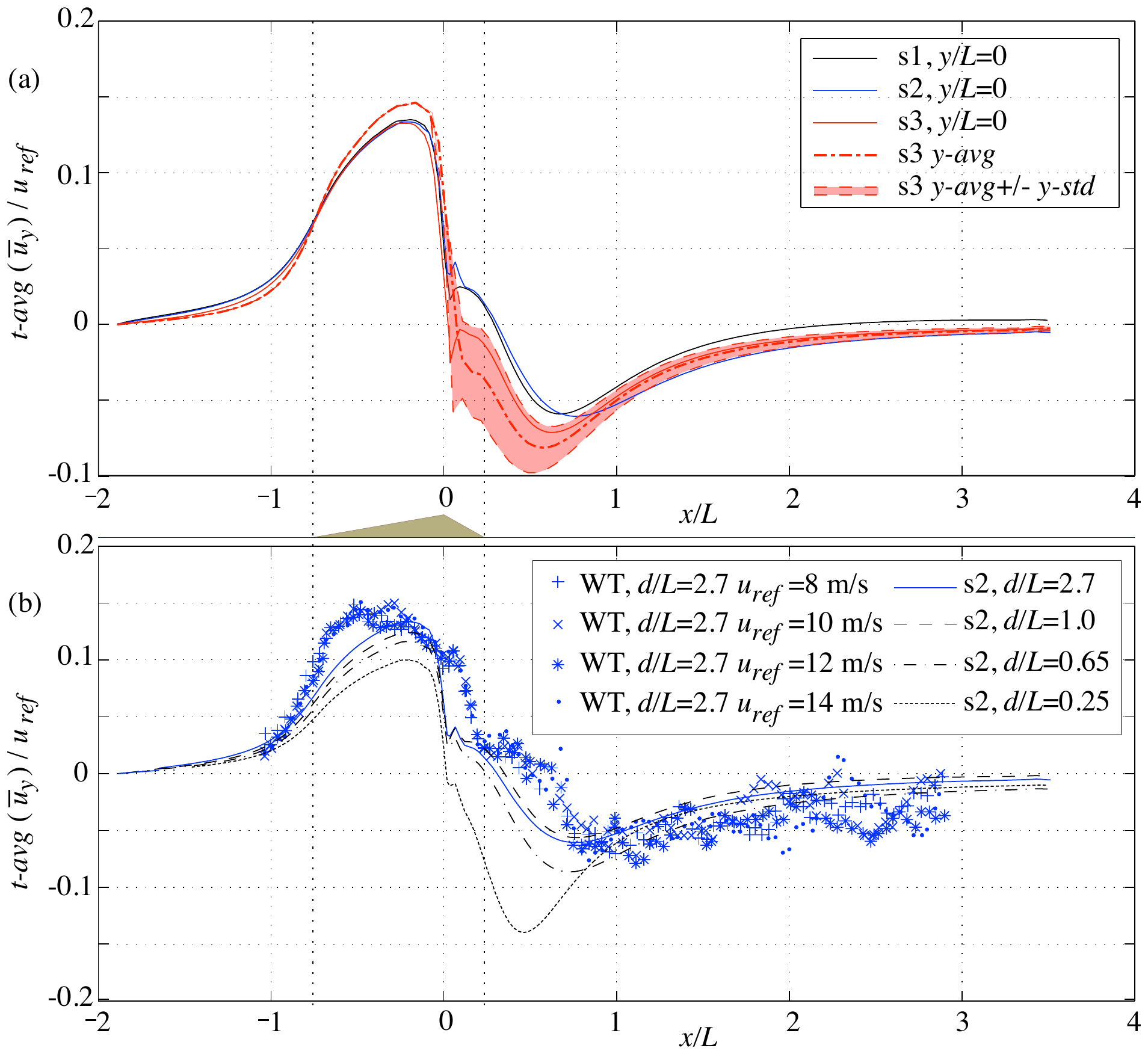}
\caption{Vertical velocity profiles: WT measurements \citep{Qian:2009dc} and present CWE results}
\label{fig:uy_x-profile_CFD_WT}
\end{center}
\end{figure}
PIV technique has been adopted on the WT vertical mid plan ($y=0$, i.e. at a distance from the side wall $d/L=2.7$) at different reference velocities of the incoming wind. The $u_y$ profiles obtained in the three computational setups are compared among them in Figure \ref{fig:uy_x-profile_CFD_WT}(a), while
the profiles measured at midspan are compared in Figure \ref{fig:uy_x-profile_CFD_WT}(b) to the computational ones at different distance from the side wall in setup s2.
Considerable differences between s1 and s2 can be observed in the far wake only, while a dramatic change in the vertical component of the velocity occurs in s3 especially in the near wake. At the dune downwind face ($0<x/h<1.7h$), $u_y$ changes its sign from positive (upward flow, s1 and s2) to negative (downward flow, s3). In other terms, the 2D clockwise recirculating region which characterizes the near wake topology in s2 at mid line no longer holds and it is suspected to change in a 3D local flow (see the large spanwise standard deviation of $u_y$).
A satisfactory agreement is observed between WT and computational approaches, if the same setup and distance from the side wall is adopted (blue points and continuous line in Fig. \ref{fig:uy_x-profile_CFD_WT}-b). Conversely, the closer the vertical plan to the side wall, the smaller the upward speed at the upwind face and the higher the downward speed at the downwind surface.
\\
An analogous post processing is proposed for the $x$-wise profiles of the shear stress magnitude $|\tau|$ at the floor in Figure \ref{fig:tau_magnitude_x-profile_CFD_WT_1}.
\begin{figure}[h]
\begin{center}
\includegraphics[width=0.75\textwidth]{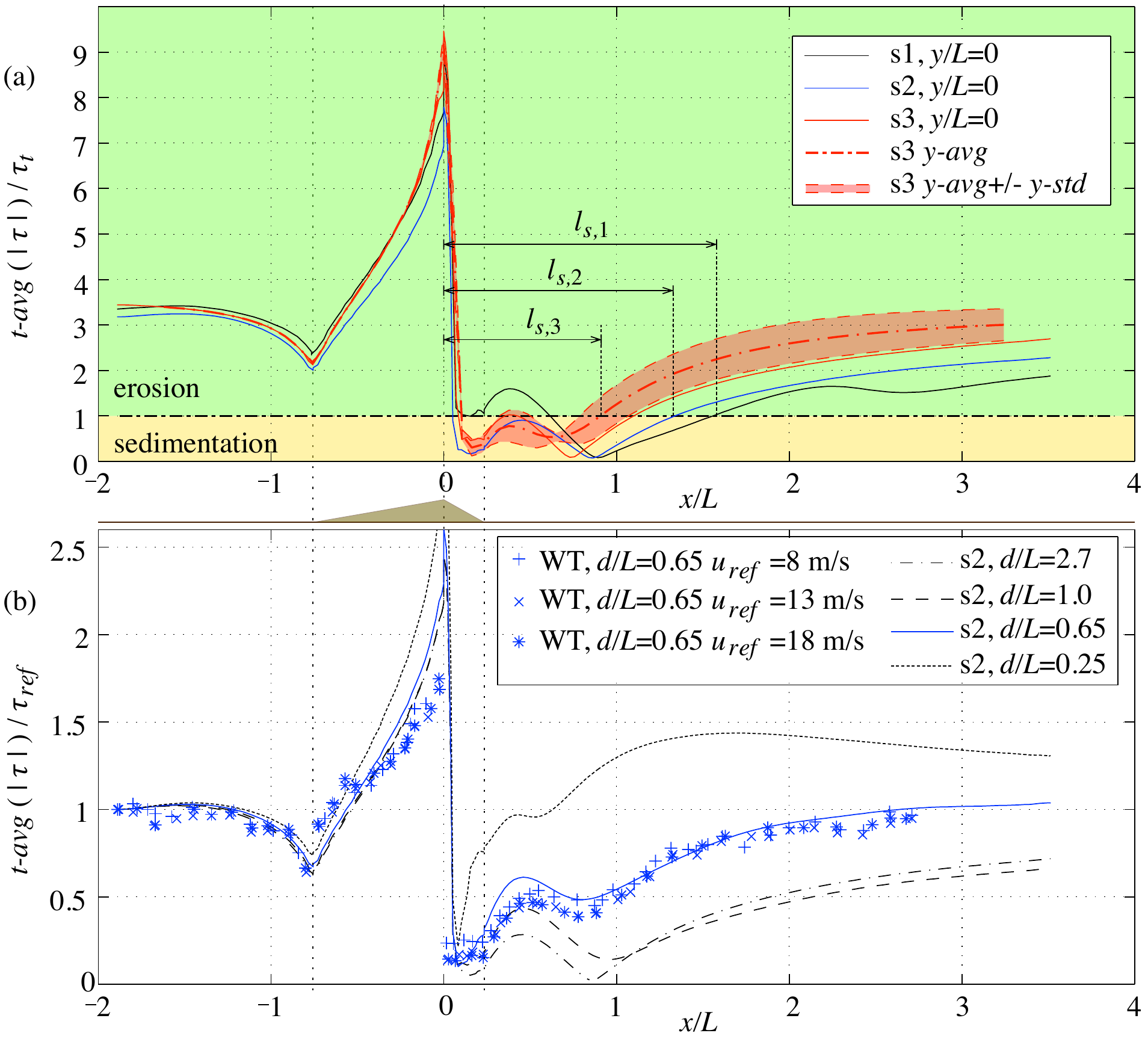}
\caption{Shear stress profiles: Wind Tunnel measurements \citep{Walker:2003dd} and present CWE results}
\label{fig:tau_magnitude_x-profile_CFD_WT_1}
\end{center}
\end{figure}
The shear stress is a quantity of particular interest in windblown sand dynamics because above a given threshold value $\tau_t$ it induces the sand grain saltation, i.e. the sand bed erosion. Conversely, at $|\tau| < \tau_t$ sedimentation of the flying grains occurs\citep{Bagnold_1941}.
For such reason, the $|\tau| / \tau_t$ obtained by computations in the three setups is plotted in Figure \ref{fig:tau_magnitude_x-profile_CFD_WT_1}(a). s1 fails in predicting the sedimentation just downwind the crest and the downwind toe. The lower the blockage ratio, the shorter the sedimentation length $l_s$, i.e.  the closer to the dune the point from which erosion takes place again. Once more, the spanwise variability of the profile along the wake is significant in s3, and the sedimentation length varies from about 0.75L to 1.1L.\\
Figure \ref{fig:tau_magnitude_x-profile_CFD_WT_1}(b) compares the shear stress magnitude profiles obtained in s2 with the one measured by \cite{Walker:2003dd} in their WT test by using Irwin-type differential pressure sensors.
{
It is worth pointing out that in this case the experimental setup differs from the computational one in the dune cross section, in the dune aspect ratio, and in the Reynolds number (see Table \ref{tab:setups}).
}
On one hand, the smaller angle of the upwind face $\alpha_u$ in WT test induces the lower shear stress magnitude along it. On the other hand, the $|\tau|$ profile downwind the crest is mainly affected by the dune aspect ratio, i.e. by the distance $d$ of the measurement alignment from the WT side wall. In fact, a very good agreement is obtained when the measurements at the WT mid plan (i.e. $d=0.5 S/L \approx 0.65$) are compared to the computational profile at the same distance from the side wall. The closer the alignment to the side wall, the higher $|\tau|$ along the wake.
In other terms, in this case study the WT wall side effects are by far larger than the ones induced by $L/h$ and/or $u_{ref}$, even if much more emphasis is traditionally given to the latter than to the former.
\subsection{3D Coherent Flow Structures}\label{sec:resu_3}
\begin{figure}[h!]
\begin{center}
\includegraphics[width=\textwidth]{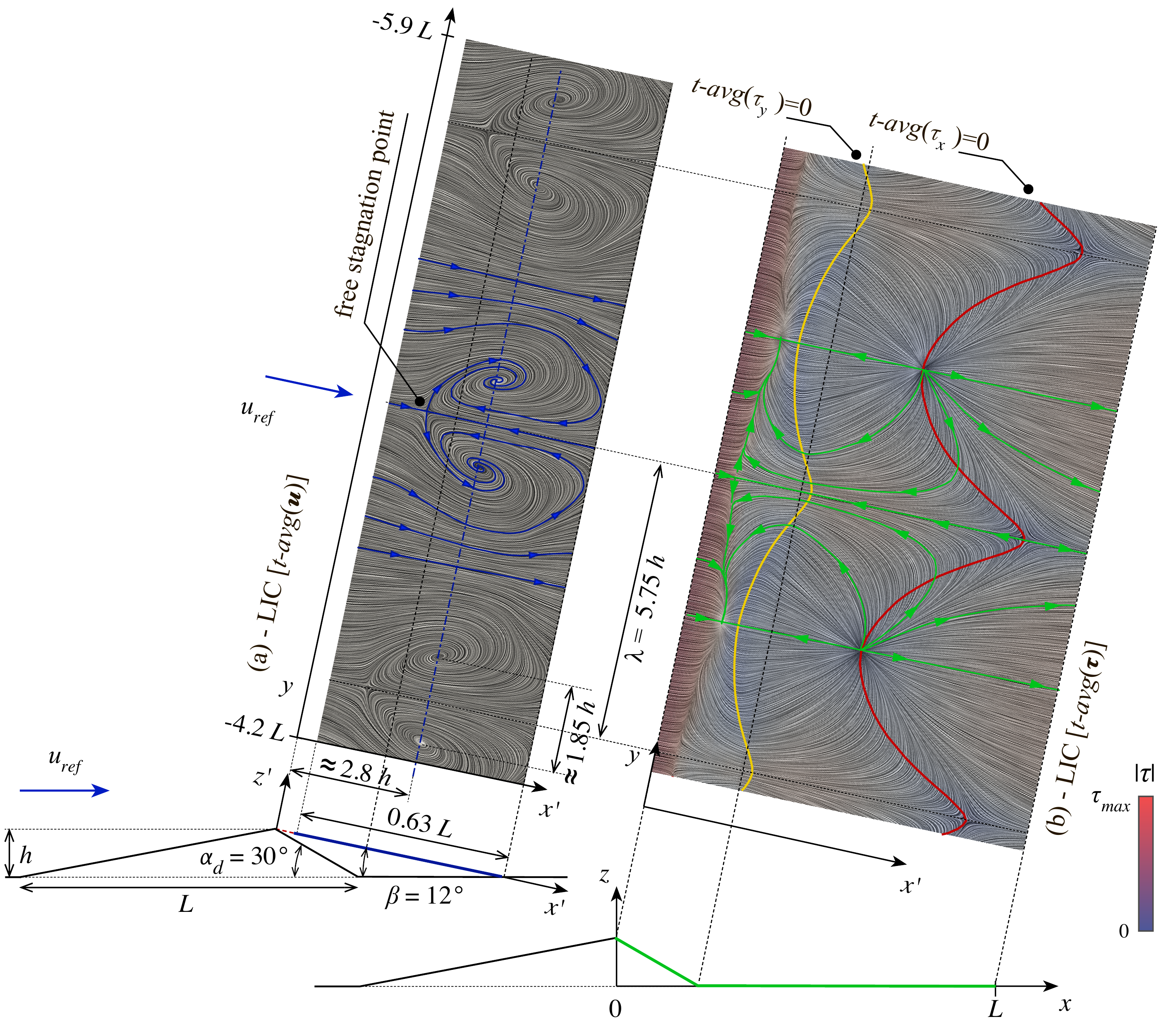}
\caption{Spanwise 3D coherent structures in the dune wake: time averaged velocity field (a), time averaged shear stress field colored by the its modulus (b)}
\label{fig:mushrooms_LIC_velo_tau}
\end{center}
\end{figure}

\begin{figure}[ht]
\begin{center}
\includegraphics[width=0.6\textwidth]{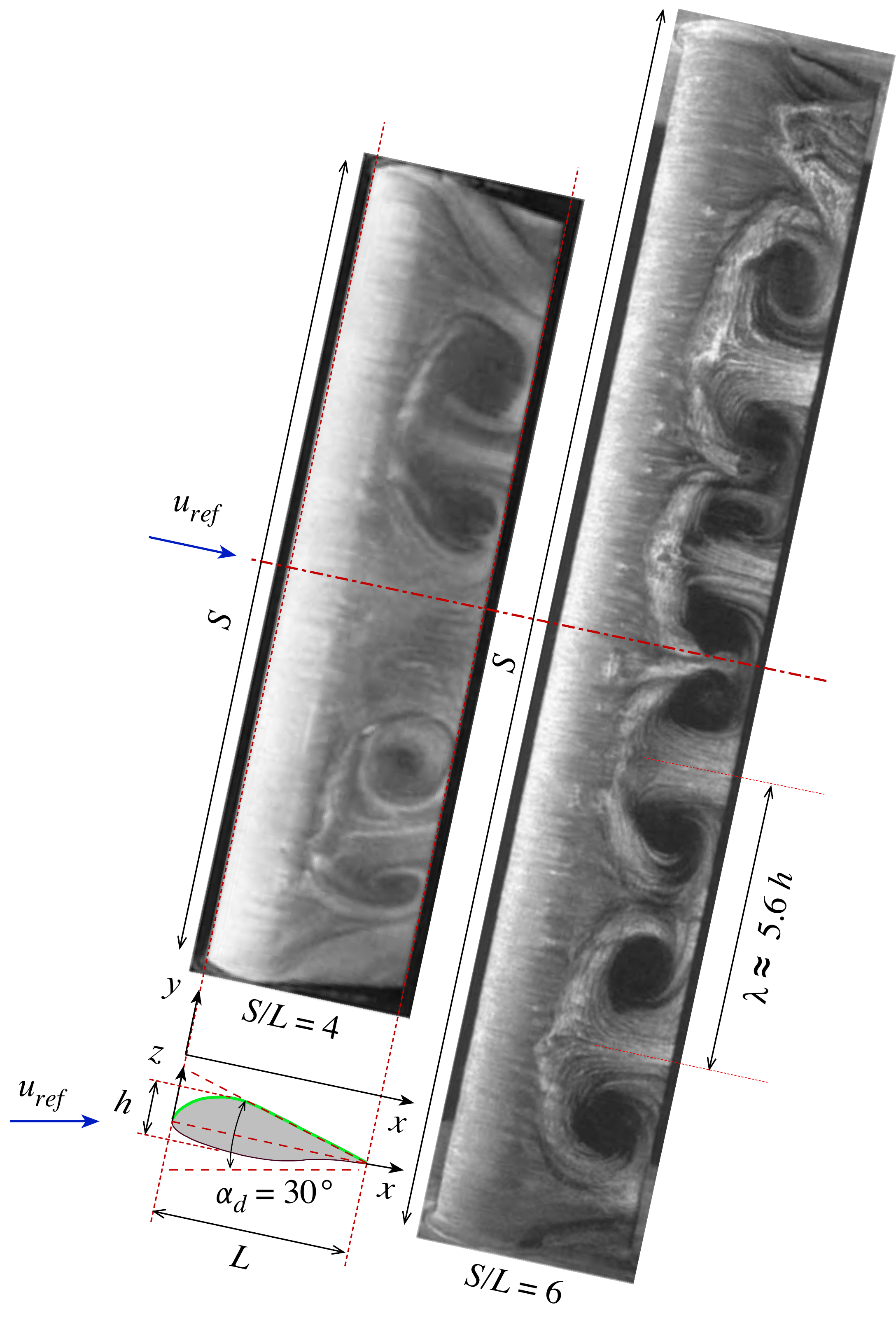}
\caption{Oil-flow pictures of the mushroom-like flow structures at the upper surface of an airfoil, after \cite{Schewe:2001p58} }
\label{fig:mush-schewe}
\end{center}
\end{figure}

This final Section aims at providing a sound phenomenological reading of the spanwise variability in simulated in external flow conditions (setup s3).
A deeper insight in the flow characteristics is then performed, thanks to the amount of information available from computational simulations. Advanced post-processing and flow visualization techniques are employed to this aim. In particular, the velocity and shear stress vector fields are visualized on selected planes by using the so-called Line Integral Convolution \citep[LIC, ][]{Cabral:1993dx, Stalling:1995bq, Laramee:2003hf}. Some information are provided herein, being a technique scarcely used in the Wind Engineering field.
\\
LIC is based upon locally filtering an input texture along a curved stream line segment in a vector field and it is able to depict directional information at high spatial resolutions.
The directional structure of a vector field can be graphically depicted by its stream lines, i.e. paths whose tangent vectors coincide with the vector field. Given a pixel $\bm{x}_0$ of the desired resulting image and numerically computed a streamline $\bm{\sigma}$, passing by $\bm{x}_0=\bm{\sigma}(s_0)$, line integral convolution consists in calculating the intensity for that pixel as
\begin{equation}\label{lic}
I(\bm{x}_0)=\int_{s_0-L}^{s_0+L}k(s-s_0)T(\bm{\sigma}(s))ds,
\end{equation}
where $s$ is the arc-length coordinate, $T$ is an input texture, $2L$ is the filter length and $k$ is a filter kernel normalized to unity. In LIC visualizations of fluid flows, a white noise or similar random image is chosen as input texture $T$. The convolution causes pixel intensities to be highly correlated along individual stream lines, but independent in directions perpendicular to them. In the resulting images the directional structure of the vector field is clearly visible.
\\
Figure \ref{fig:mushrooms_LIC_velo_tau}(a) and (b) show the LIC visualization of the time-averaged velocity field on an ideal inclined plane ($\beta=12^{\circ}$) crossing the near wake from the dune crest, and the LIC visualization of the time-averaged shear stress field on the dune downwind face and wake floor, respectively.
{In both figures, the direction of the filed is highlighted by superimposed arrowed lines.}
The most striking result is that in Figure \ref{fig:mushrooms_LIC_velo_tau}(a), where the LIC of the time-averaged velocity field highlights quasi-steady, "mushroom"-like coherent flow structures having a pair of owl eyes each. These structures consists of two counterrotating foci with a free stagnation point downwind the crest and between the foci. The main dimensions of such structures are given in the figure.
The corresponding LIC of time-averaged shear stress field (Fig. \ref{fig:mushrooms_LIC_velo_tau}-b) is even more complicated and characterized by a "skein"-like pattern.
{
The pattern shape appears to be strongly related with the mushroom-like structure, especially at the ground floor (i.e. where the mushroom is closest to the floor).
In order to shed some more light in such a pattern, it is colored by the modulus of the same shear stress, and the spanwise lines along which $\tau_y=0$ and $\tau_x=0$ (reattachment line, see also Fig. \ref{fig:reattachment}) are superimposed. For both lines a geometrical reading with respect to the LIC pattern can be provided: the former is the locus of points where the tangent to the streamlines are aligned with the $x$-axis, while the latter is the locus of points where the tangent to the streamlines are aligned with the $y$-axis. 
The mushroom spanwise characteristic length is equal to the wavelength $\lambda$ of the "festoon" reattachment line. In particular, the spanwise maximum reattachment length correspond to the mushroom center, while the minimum value takes place in between two consecutive mushrooms.
}
\\
To the authors' best knowledge, analogous spanwise coherent flow structures have been previously recognized at least in two studies involving nominally 2D setups.
\\ 
Similar spanwise mushroom-shaped flow structures have been visualized by
\cite{Schewe:2001p58} around an apparently different aerodynamic setup, i.e. the separated flow around an airfoil at incidence in the critical regime. In fact, in his pioneering and seminal work Schewe has observed and described in detail mushroom-like flow structures (Figure \ref{fig:mush-schewe}), combined in different arrangements for different values of the airfoil aspect ratio $S/L$ ($L=w$), and bounded by disturbed regions close to the end plates.
On the basis of the experimental results, Schewe conjectures that: 
i. the occurrence and number of the mushroom-like structures depend on the inclination of the base upper surface;
ii. the spanwise flow structure remains intact as $S/L \to \infty$;
iii. the wavelength $\lambda$ scales with the sectional dimension(s) of the airfoil;
iv. preferential or more natural spanwise arrangements exist, in terms of number of cells and their wavelength. These speculations seem to hold{, to a certain extent, }
 also in the present case study.
 {
 In fact:
 i. the downwind face of the dune and the base upper surface of the airfoil share the same inclination with respect to the incoming wind ($\alpha_d \approx 30 ^\circ$);
 ii. setup s3 confirms that the coherent flow structure is qualitatively unchanged by passing from s3a to s3b, where the latter mimics an infinite aspect ratio of the dune because of the periodic side b.c. and the increased spanwise length;
 iii. the characteristic length $\lambda$ of the mushrooms found by \cite{Schewe:2001p58} is comparable to the one in s3, if both are scaled versus $h$ ($\lambda \approx 5.75h$, Fig. \ref{fig:mushrooms_LIC_velo_tau}-a, $\lambda \approx 5.6h$, Fig. \ref{fig:mush-schewe});
 iv. in both Schewe and present setups, stable arrangements are characterized by a given number of spanwise aligned mushrooms. In Schewe, an even number of vortices is always observed, but the end-plate disturbances considerably extend spanwise; conversely, in s3 an odd number of mushrooms emerges in both s3a (7) and s3b (27) setups (see Fig. \ref{fig:reattachment}), despite the spanwise length in s3b is four time the one in s3a. It is worth pointing out that the quasi-periodic mushroom structure is corrupted at midspan, if the dune length $S=w$ is not an odd multiple of $\lambda$ (s3b).
}
Despite the analogies above, it is worth pointing out that the flow structures in \cite{Schewe:2001p58} are recognized only in the critical regime, while the occurrence of the same condition is unlikely in the present case study.
\\
More recently, analogous 3D, global, quasi-steady vortices have been simulated by \cite{spalart_2014}  in an even more different fluid flow setup, i.e. the nominally 2D, high Reynolds number, fully developed turbulent Couette flow. The RANS-based simulations of \cite{spalart_2014} (including the SST $k-\omega$ one), successfully reproduce the same flow structures previously observed and simulated in experiments and DNS, respectively. In the present perspective, it is worth pointing out that:
i. the vortices are arranged in a quasi-periodic z-wise system of counter-rotating pairs (Fig. 1, \cite{spalart_2014}), analogously to the arrangement in Figure \ref{fig:mushrooms_LIC_velo_tau}(a) ;
ii. the distribution of the skin-friction coefficient in the $z$ direction (Fig. 6, \cite{spalart_2014}) qualitatively matches the "festoon-shaped" spanwise reattachment line in Figure \ref{fig:reattachment};
iii. the simulated flow structures hold in a wide range of Reynolds numbers, and the effect of the vortices is only weakly dependent on Re.
In their outlook, Spalart et al remark that there is a convincing scale separation between the simulated flow structures and the turbulence which is represented by the RANS model. The vortices have a lateral scale much smaller than the scale of the geometry (which is infinite in the setup), and random locations in $z$, giving them the nature of an instability. 
The remarks above qualitatively hold also in the present case study.
Despite such analogies, it is worth pointing out that the flow structures in \cite{spalart_2014} are somewhat triggered by an initial velocity field mimicking a periodic system of counter-rotating vortex pairs in the ($y-z$) plane, while in the present case study the mushroom-like structures spontaneously develop from uniform initial conditions.
\\
%
%
\section{Conclusions}
%
The present study points out some emerging 3D coherent flow structures in the wake of a transverse dune under different setup conditions by means of computational simulations and compares the obtained results with a number of experimental wind tunnel measurements available in literature. The comparison shows a good and robust agreement of the CWE results to WT data, and it goes beyond: it puts in evidence common misunderstandings in setting up computational and experimental models for CWE/WT validation.
Some good practices in dune aerodynamics CWE simulations and WT tests are recommended: i. CWE-WT comparisons require the careful simulation of the WT setup, including the wind tunnel geometry; ii. forcing the flow symmetry by b.c. in CWE can strongly affect the results; iii. high values aspect ratio are recommended in WT test to replicate external flow conditions, e.g. $w/L \gg 10$.\\
Surprisingly, emerging mushroom-like, coherent flow structures in the dune wake are clearly shown by a deep analysis of the CWE results. They are compared to analogous flow structures arising from rather different aerodynamic setups belonging to referential literature in fluid dynamics.
It follows that the studied setup can be ascribed to a wider class of 3D flows occurring under 2D nominal conditions. According to the authors such class of flow seems to be of high relevance in modern developments of fluid dynamics, and it still remain poorly understood in its general features, if any. The authors hope further independent studies will appear and further develop their contribution.
In particular, the precise definition of the aerodynamic regime(s) at which mushrooms grow in the dune near wake remain an open issue. An even wider computational study would be required to discuss the permanence and/or changes of such coherent flow structures form the model scale to the full scale, e.g. by varying Reynolds number and/or surface roughness.
\section*{Acknowledgments}
The study has been developed in the framework of the Windblown Sand Modeling and Mitigation joint research, development and consulting group established between Politecnico di Torino and Optiflow Company. The authors wish to thank Luigi Preziosi and Nicolas Coste, members of the WSMM group, for the helpful discussions about the topics of the paper.
G{\"u}nter Schewe is gratefully acknowledged for his kind availability in providing his flow visualizations and for his stimulating remarks.
\bibliography{biblio-JWEIA_ANIV_2014_Bruno_Fransos-doi}
\end{document}